\documentclass[aps,prl,twocolumn,nofootinbib,superscriptaddress]{revtex4-2}
\usepackage[utf8]{inputenc}
\usepackage{amsmath,amssymb,physics}
\usepackage{graphicx}
\usepackage{xcolor}
\usepackage{braket}
\usepackage{hyperref}
\hypersetup{colorlinks=true, linkcolor=blue, citecolor=blue, urlcolor=blue}
\usepackage{array}
\usepackage{svg}
\usepackage{appendix}
\usepackage{float}
\usepackage{bbold}

\newcommand{\mmn}{\vspace{0.09in}\noindent}

\begin{document}

\title{Realistic Gottesman-Kitaev-Preskill stabilizer states enable universal quantum computation}

\author{Fariba Hosseinynejad}
\affiliation{Institute for Quantum Science and Technology, and Department of Physics and Astronomy,
University of Calgary, Calgary, Alberta, Canada T2N 1N4}

\author{Pavithran Iyer}
\affiliation{Xanadu Quantum Technologies Inc., Toronto, Ontario, Canada M5G 2C8}
\author{Guillaume Dauphinais}
\affiliation{Xanadu Quantum Technologies Inc., Toronto, Ontario, Canada M5G 2C8}
\date{\today}
\author{David L. Feder}
\affiliation{Institute for Quantum Science and Technology, and Department of Physics and Astronomy,
University of Calgary, Calgary, Alberta, Canada T2N 1N4}

\date{\today}

\begin{abstract}
Physical Gottesman-Kitaev-Preskill (GKP) states are inherently noisy as ideal ones would require infinite energy. While this is typically considered as a deficiency to be actively corrected, this work demonstrates that imperfect GKP stabilizer states can be leveraged in order to apply non-Clifford gates using only linear optical elements. In particular, Gaussian operations on normalizable GKP states, combined
with homodyne measurements, permit two key primitives: clean projection onto 
Pauli eigenstates in the normalizable GKP codespace, thereby implementing 
Clifford gates with high fidelity; and probabilistic projection of unmeasured 
modes onto non-Pauli eigenstates. These results demonstrate that normalizable
GKP stabilizer states combined with Gaussian operations provide a practical framework for 
computational universality within the measurement-based model of quantum 
computation in a realistic continuous-variable setting.
\end{abstract}

\maketitle

\noindent \textit{Introduction}---
The Gottesman--Kitaev--Preskill (GKP) codes~\cite{gottesman2001encoding} was 
introduced more than two decades ago, but has recently received renewed 
attention due to significant experimental progress in preparing, manipulating,
and measuring GKP states~\cite{nature2025,Xanadu2025Modular,Matsos2025,Brady2024,bourassa2021blueprint,Campagne-Ibarcq2020-gt,Eickbusch_2022,Sivak_2023,Kudra_2022,PhysRevLett.132.150607,science.adk7560,Fl_hmann_2019,De_Neeve2022-wv}. The 
central idea is to encode a discrete-variable (DV) qubit into the 
continuous-variable (CV) position and momentum degrees of freedom of a harmonic 
oscillator, with the dual motivation of achieving intrinsic noise resilience 
and enabling universal quantum computation within CV architectures. Since 
Gaussian states combined with Gaussian operations are efficiently classically 
simulatable and susceptible to noise~\cite{bartlett2002efficient}, GKP 
codes---being inherently non-Gaussian---provide a natural candidate for 
enabling fault-tolerant computational universality.

In the ideal square single-mode GKP encoding, logical $\ket{0}$ and $\ket{1}$ 
registers correspond 
to an infinite superposition of infinitely squeezed position eigenstates with 
peaks spaced by $\sqrt{\pi}$~\cite{gottesman2001encoding}, in units where 
$\hbar = 1$. However, these 
states are unphysical, as they require infinite energy. Realistic GKP states 
are instead constructed by introducing a Gaussian envelope over the grid peaks, 
effectively damping the high-photon-number components and making the states 
normalizable~\cite{menicucci2014fault,gottesman2001encoding}. We refer to these 
as \emph{Fock-damped GKP states}, characterized by a single parameter 
controlling both the envelope and peaks width~\cite{Grimsmo2021GKP}. The 
challenge is to perform logical operations on quantum states encoded in
Fock-damped GKP registers. In the ideal case, all Clifford gates correspond to 
Gaussian operations~\cite{gottesman2001encoding}; for example, a logical 
Pauli-$X$ is simply a position quadrature displacement. However, na\" ive 
Gaussian displacements on realistic GKP states distort the envelope and reduce 
fidelity, requiring a more careful treatment of logical gate 
implementation~\cite{Tzitrin2020}. Furthermore, to 
achieve universality, access to at least one non-Clifford element -- either a 
gate or an ancillary state -- is necessary.

A powerful alternative to the circuit model is \emph{measurement-based quantum 
computation} (MBQC), developed in both DV~\cite{raussendorf2001one} and 
CV~\cite{menicucci2006universal} settings. MBQC replaces explicit gate 
sequences with adaptive local measurements on a highly entangled resource 
state, for example the cluster 
state~\cite{Asavanant2019Time,Larsen2019Deterministic,Yokoyama2013Ultra}. 
Measurement-based gate teleportation is also a natural strategy for 
implementing both Clifford and non-Clifford operations in multimode entangled
GKP states; however, previous analyses of such 
protocols~\cite{walshe2020continuous} assumed ideal GKP states. In that case,
only single-qubit Clifford gates can be teleported: the unmeasured modes always 
project onto Pauli eigenstates, rendering the circuit non-universal. 

More recently, it was observed that introducing certain ancillary Gaussian states could in principle supply the missing resourcefulness~\cite{All-Gaussian,calcluth2023vacuum}, but a complete understanding of how to practically and reliably teleport both Clifford and non-Clifford gates in a realistic setting is missing. In particular, Ref.~\cite{All-Gaussian} employ ideal GKP Pauli eigenstates and requires two rounds of GKP error correction on a Gaussian input, producing only one of twelve Clifford-equivalent $H$- or $T$-type magic states (rather than arbitrary states on the Bloch sphere) with less than a 10\% success probability per attempt. Ref.~\cite{calcluth2023vacuum} shows that circuits composed of \emph{ideal} GKP stabilizer states, Gaussian operations, and homodyne detection are efficiently classically simulatable, and identifies the vacuum as an additional Gaussian resource that can break this simulatability. This leaves a clear knowledge gap: a resource-efficient protocol employing only realistic GKP states that enables higher-probability teleportation of genuinely non-Clifford gates, allows access to arbitrary non-Pauli states on the Bloch sphere, and avoids multiple rounds of experimentally costly GKP error correction.

A related literature has focused on identifying the resources required for 
universality in hybrid CV-DV settings, where the physical system is 
continuous-variable but the logical layer depends on access to non-Clifford 
gates~\cite{bartlett2002efficient,Davis2024}. Several works have introduced 
CV-to-DV mapping techniques~\cite{Shaw2024,Pantaleoni2020}, combined with DV 
magic monotones to quantify resourcefulness~\cite{Calcluth2024,Davis2024}, 
showing that circuits of ideal GKP states, Gaussian operations, and homodyne 
detection are efficiently classically 
simulatable~\cite{Calcluth2022efficient,calcluth2023vacuum,All-Gaussian}. 
However, these analyses fundamentally rely on ideal GKP states and defer a
fully realistic treatment due to the difficulty of simulating and analytically 
characterizing finite-squeezing 
effects~\cite{calcluth2023vacuum,Calcluth2022efficient,Calcluth2024}.

In this work, we settle the open question of whether realistic, Fock-damped GKP 
states can enable universality by proposing a minimal, resource-efficient 
gate-teleportation protocol that works directly in the CV domain without 
invoking standard CV-to-DV mappings to quantify resourcefulness. The protocol 
uses a simple Gaussian circuit: two squeezed and Fock-damped GKP \(\ket{+}\) 
states interfere on a balanced beamsplitter, followed by a phase shifter and 
a q-homodyne measurement. The output-state probability distribution reveals a 
crucial dichotomy: (1) for ideal GKP states, the protocol is limited to 
deterministically teleporting Clifford gates, as the unmeasured mode always 
projects onto a Pauli eigenstate, whereas (2) for realistic, Fock-damped GKP
states, tuning the damping parameter and phase angle enables high-probability 
projections onto both Pauli and non-Pauli eigenstates, including
magic-state-like outputs, with success probabilities $\sim 0.4$ for fidelities 
$F \gtrsim 0.96$ using experimentally relevant squeezing. We thereby establish 
that the finite-energy envelope of realistic GKP states, modeled here via Fock 
damping, becomes a resource for universal quantum computation instead
of a liability. 

\noindent \textit{GKP Encoding}---
GKP registers consist of a one-dimensional comb 
$\ket{s}_L = \sum_{j\in\mathbb{Z}} \ket{(2j+s)\sqrt{\pi}}_q$, $s=0,1$, of
oscillators expressed in the position quadrature $\hat{q}=\frac{1}{2}\left(
\hat{a}+\hat{a}^{\dag}\right)$, where $\hat{a}$ ($\hat{a}^{\dag}$) are photon 
annihilation (creation) operators (the momentum quadrature is 
$\hat{p}=\frac{i}{2}\left(\hat{a}-\hat{a}^{\dag}\right)$ such that the 
commutator is $[\hat{q},\hat{p}]=i$). Position eigenstates are 
$|x\rangle_q = \sum_{n=0}^{\infty}\psi_n(x)|n\rangle$ and $\psi_n(x) 
= \frac{1}{\sqrt{2^n n!}}\left(\frac{1}{\pi}\right)^{1/4} e^{-x^2/2}H_n(x)$ are 
harmonic oscillator wave functions with $H_n(x)$ `physicist' Hermite 
polynomials. The GKP states are non-normalizable and therefore 
unphysical. Realistic GKP states $\ket{\tilde{s}}_L=e^{-\beta\hat{n}}\ket{s}_L$
are finitely squeezed and Fock-damped with strength $\beta$ via a Gaussian 
operator $\mathcal{N}_\beta=e^{-\beta\hat{n}}$, where $\hat{n}$ is the number 
(Fock) operator $\hat{n}=\hat{a}^{\dag}\hat{a}$; this yields logical states
$\ket{\tilde{s}}_L= \sum_{j\in\mathbb{Z}}\sum_{n=0}^{\infty}\psi_n((2j+s)
\sqrt{\pi}) e^{-\beta n}\ket{n}$. Throughout the remainder of this work, the 
logical subscript \(L\) is omitted for notational simplicity. Unless otherwise 
stated, all references to states such as \(|0\rangle\) and \(|1\rangle\)
refer to their respective ideal GKP states; a tilde denotes Fock-damped 
normalizable physical GKP states (e.g.\ \(\ket{\tilde{0}}\)).
Physically, $\beta$ is set by the quality of the state-preparation
process, and is not a dynamically adjustable parameter.  Different values of $\beta$ correspond to prepared GKP resource states with
different intrinsic quality.
It is important to stress that the Fock-damping operator
$e^{-\beta\hat{n}}$ is used here as a convenient single-parameter model
for finite-energy approximate GKP states, following
Ref.~\cite{Grimsmo2021GKP}.  Realistic preparation schemes in
superconducting circuits, trapped ions, and photonics generate grid
states whose Fock-space populations decay approximately as
$\propto e^{-\beta n}$; the family
$\ket{\tilde{s}} = e^{-\beta\hat{n}}\ket{s}$ provides an effective
description of these physical states that can be approached ``from
below'' by increasing the number of superposed components and the
available squeezing.  
The Fock damping operator $e^{-\beta\hat{n}}$ is not a dynamical gate within the protocol.

\begin{figure}[t]
\centering
\includegraphics[width=\linewidth]{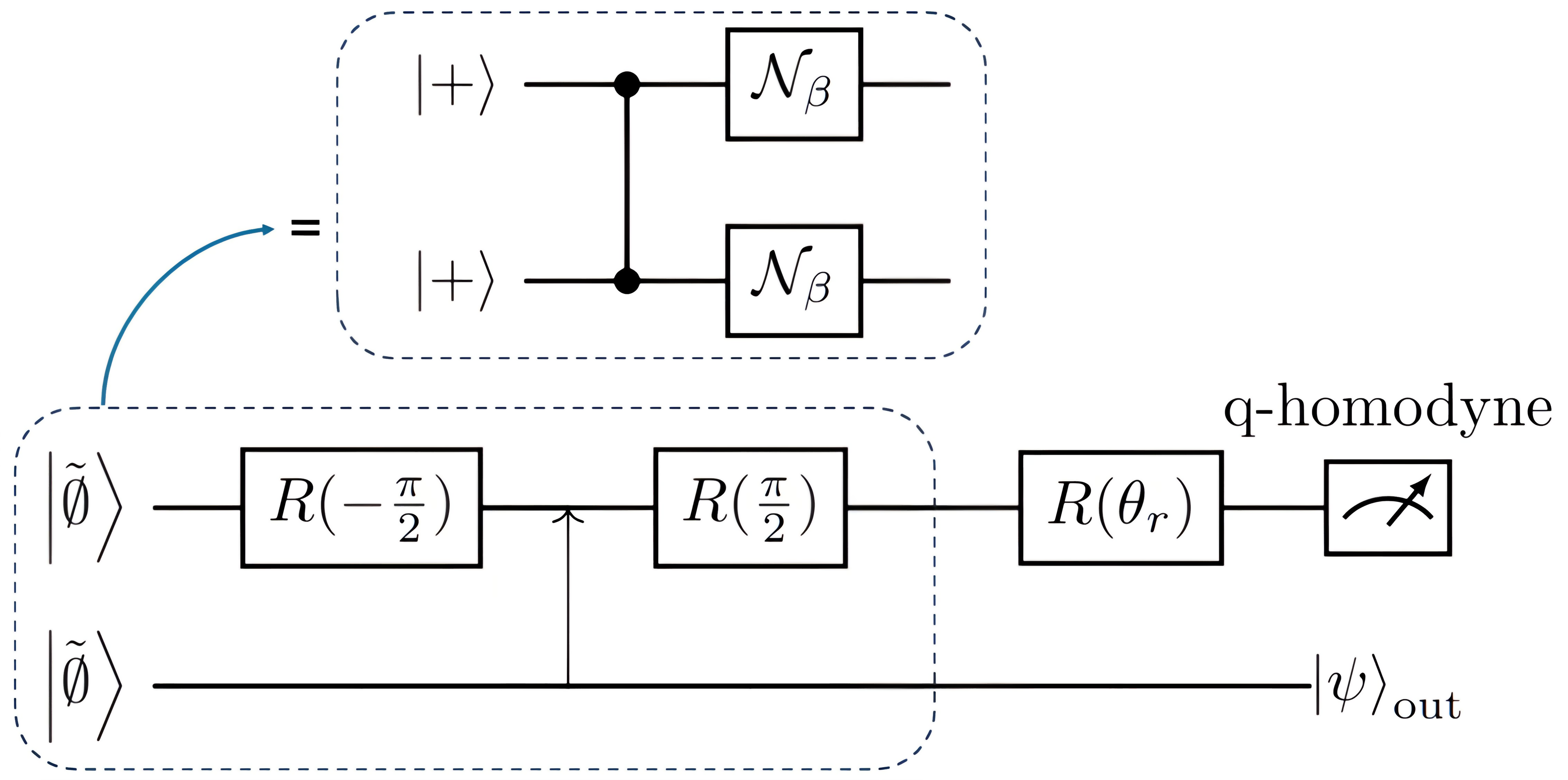}
\caption{Quantum circuit for gate teleportation}
\label{fig:circuit}
\end{figure}

\noindent \textit{Teleportation Circuit and Output State}---
Consider two modes $a=1,2$, each prepared in Fock-damped `sensor states'
$|\tilde{\emptyset}\rangle=\mathcal{N}_\beta|\emptyset\rangle$, where
$|\emptyset\rangle=\sum_{j\in\mathbb{Z}}\ket{j\sqrt{2\pi}}_q$. These states can
be subsequently entangled via a 50-50 beamsplitter $BS$, depicted as an arrow 
pointing from one mode to the other in Fig.~\ref{fig:circuit}, which maps
$\{\hat{q}_a,\hat{p}_a\}\to\frac{1}{\sqrt{2}}\{\hat{q}_1+(-1)^a\hat{q}_2,
\hat{p}_1+(-1)^a\hat{p}_2\}$. In practice, it is more convenient to conjugate
the beamsplitter by a rotation on mode 1, so that the input state 
becomes~\cite{walshe2020continuous,PRXQuantum.2.040353,PhysRevA.110.012436,nature2025}
\begin{eqnarray}
|\Psi\rangle_{\rm in}&=&R_{-\pi/2,1} BS_{1,2}R_{\pi/2.1}\ket{\tilde{\emptyset}}
\ket{\tilde{\emptyset}}\nonumber \\
&=&\mathcal{N}_{\beta,1}\mathcal{N}_{\beta,2}R_{-\pi/2,1} BS_{1,2}R_{\pi/2.1}
\ket{\emptyset}\ket{\emptyset}\nonumber \\
&=&\mathcal{N}_{\beta,1}\mathcal{N}_{\beta,2}CZ_{1,2}|+\rangle|+\rangle,
\end{eqnarray}
where $CZ_{1,2} = e^{-i \hat q_1 \hat q_2}$ is the controlled-phase gate, the 
rotation operator $R_{\theta,a}=e^{i\theta\hat{n}_a}$, and 
$|+\rangle=\frac{1}{\sqrt{2}}\left(|0\rangle+|1\rangle\right)$. Note that the 
beamsplitter operator preserves the total photon number, and therefore commutes
with damping operators on both modes as well as phase shifters. Thus, even
though the input states are Fock-damped sensor states, the preparation circuit
is mathematically equivalent to Fock-damping both modes of a cluster state of 
formed of entangled GKP states.

In DV case, measurement-based gate teleportation is accomplished by applying a 
rotation to the first qubit, followed by a computational-basis measurement; in 
contrast, in the CV case the rotation $R_{\theta_r}$ is applied to the first
mode in quadrature (not logical) space. Applying this rotation and performing 
a $q$-homodyne measurement on the first mode with outcome $q_m$, as depicted in
Fig.~\ref{fig:circuit}, one obtains (ignoring normalization) 
$\ket{\Psi}_{\text{out}} =C_+\,\ket{\tilde{+}}+C_-\,\ket{\tilde{-}}$ on the 
second mode, where $C_+=\cos(\theta/2)$ and $C_-=e^{i\phi}\sin(\theta/2)$ are 
expressed in terms of the angles defining the Bloch sphere. The coefficients 
$C_{\pm}$ are derived in the Appendix: 
\begin{equation}
C_{\pm}=\frac{e^{-i\frac{\pi}{2}k^2\cot\zeta}}
{\sqrt{1-e^{2i\zeta}}}\sum_{j=-\infty}^{\infty}e^{-i\frac{\pi}{2}(2j+s)^2
\cot\zeta}e^{i\pi(2j+s)k\csc\zeta},
\label{eq:Cplusminus}
\end{equation}
where $s=0,1$ for $C_+,C_-$, respectively; here $\zeta\equiv\theta_r+i\beta$
and $k\equiv q_m/\sqrt{\pi}$ is a dimensionless parameter that rescales the 
homodyne measurement outcome $q_m$ in natural GKP grid units. The coefficients 
therefore depend on the measurement outcome $q_m\to k$ as well as the 
adjustable parameters $\theta_r$ and $\beta$.

The coefficients can be equivalently expressed as 
\begin{equation}
C_{\pm}=\frac{e^{i\frac{\pi}{2}k^2\tan\zeta}}
{\sqrt{2\left(1+e^{2i\zeta}\right)}}\vartheta_3\left\{
-\frac{k\pi}{2}\sec\zeta+\frac{\pi s}{2},e^{i\frac{\pi}{2}\tan\zeta}\right\}.
\end{equation}
Apart from a common multiplicative factor, which is not merely a global phase 
since $\zeta$ is complex, the coefficients are proportional to Jacobi theta 
functions of the third and fourth kinds:
\begin{eqnarray}
C_+&\propto&\vartheta_3\left\{z,\omega\right\}=\vartheta_3\left\{z|\tau\right\};
\nonumber \\
C_-&\propto&\vartheta_3\left\{z+\frac{\pi}{2},\omega\right\}
=\vartheta_4\left\{z,\omega\right\}=\vartheta_4\left\{z|\tau\right\},
\end{eqnarray}
where $z=-\frac{\pi}{2}k\,\sec\zeta$ and the nome which is usually represented 
as $q$ is here denoted as $\omega=e^{\frac{i}{2}\pi\tan\zeta}
\equiv e^{i\pi\tau}$ to prevent confusion with the position value; note that 
$\tau=\frac{1}{2}\tan\zeta$. The Jacobi theta functions are defined as
\begin{equation}
\vartheta_3\{z,\omega\}=\sum_{\ell=-\infty}^{\infty} e^{2i\ell z}
\omega^{\ell^2};\;\vartheta_4\{z,\omega\}=\vartheta_3\left\{z+\frac{\pi}{2}
,\omega\right\}.
\label{thetafunction}
\end{equation}

For small but non-zero damping, $\beta\ll 1$, one may expand 
$\tan\zeta\approx\tan\theta_r+i\beta\sec^2\theta_r$ to lowest order in $\beta$, 
such that
\begin{equation}
\vartheta_3\{z,\omega\}=\sum_{\ell}e^{-\pi\beta\sec^2\theta_r\ell^2/2}
e^{i\pi\tan\theta_r\ell^2/2}e^{-2i\ell z}.
\end{equation}
Consider the case where the applied rotation angle is obtained via the rational 
approximation $\tan\theta_r=u/v\in\mathbb{Q}$. In any practical 
setting, one can always replace $\tan\theta_r$ by a rational value consistent 
with its value to experimental precision, so this approximation is not overly 
restrictive. Then the summation above can be decomposed as $\ell=2vm+n$, where 
$m,n\in\mathbb{Z}$ and $n=0,1,\dots,2v-1$. Defining the effective damping 
parameter $\beta':=\frac{1}{2}\left(1+\frac{u^2}{v^2}\right)\beta$, after some 
lengthy algebra one obtains
\begin{align}
\vartheta_3\{z,\omega\} &= \frac{1}{2v\sqrt{\beta'}} \sum_{n=0}^{2v-1} e^{i\pi \frac{n^2 u}{2v}} \notag\\
&\quad \times \sum_{\ell=-\infty}^{+\infty} e^{i\pi \ell n / v} 
\exp\left[-\frac{(z + \ell \pi / 2v)^2}{\pi \beta'}\right],
\label{eq:theta_rational}
\end{align}
and the expression for $\theta_4(z,\omega)$ is the same other than an 
additional factor of $(-1)^n$ in the first sum.

\noindent \textit{Only Pauli Eigenstates Result for Zero Damping}---
For $\beta\to 0$ the post-measurement outcome is a Pauli eigenstate for any 
choice of $\theta_r$. The results are sketched here, and additional details are 
provided in the Appendix. 
In the limit of zero damping, each 
Gaussian term appearing in the sum over $\ell$ in Eq.~(\ref{eq:theta_rational})
becomes a Dirac delta function; using the identity
\begin{equation}
\delta(z) = \lim_{\beta \to 0} \frac{1}{\pi\sqrt{\beta}}
\exp\left(-\frac{z^2}{\pi\beta}\right),
\end{equation}
one obtains
\begin{align}
\vartheta_3\{z,\omega\} &= \frac{\pi}{2v} \sum_{n=0}^{2v-1} e^{i\pi \frac{n^2 u}{2v}}
\sum_{\ell=-\infty}^{+\infty} e^{i\pi \ell n / v}
\delta\left(z + \frac{\ell \pi}{2v}\right).
\end{align}
Therefore, the coefficients of the output state $C_{\pm}$ are non-zero only
when $z=-\frac{\ell\pi}{2v}$. In the zero-damping limit,
$z\to -\pi k\sec\theta_r/2$ and so only the values $q_m=k\sqrt{\pi}$ that will 
be obtained correspond to $k = \frac{\ell}{\sqrt{u^2 + v^2}}:=k_m$, 
$\ell\in\mathbb{Z}$. In this limit, 
\begin{equation}
\frac{C_-}{C_+}=
\left.\sum_{n=0}^{2v-1} e^{i\pi \frac{n^2 u}{2v}}e^{i\pi(\ell+v)n/v}\middle /
\sum_{n=0}^{2v-1} e^{i\pi \frac{n^2 u}{2v}}e^{i\pi\ell n/v}\right. ,
\end{equation}
i.e.\ the ratio of generalized quadratic Gauss sums
\begin{equation}
G(a,b,c)=\sum_{n=0}^{c-1}\exp\left(2\pi i\frac{a n^2 + b n}{c}\right).
\end{equation}
Using the known properties of these sums, it is straightforward to prove 
that the ratio \(C_-/C_+\) is restricted to \(\pm 1\), \(\pm i\), or 
$\{0,\infty\}$, i.e.\ that the output always corresponds to Pauli eigenstates
of undamped GKP basis states. In particular: if $u$ and $v$ are both odd, then 
the output is a Pauli-Y eigenstate; if $u$ is even and $v$ is odd then one 
obtains an X eigenstate; and if $u$ is odd and $v$ is even then the output is a 
Z eigenstate.

\begin{figure*}[t]
\includegraphics[width=1\linewidth]{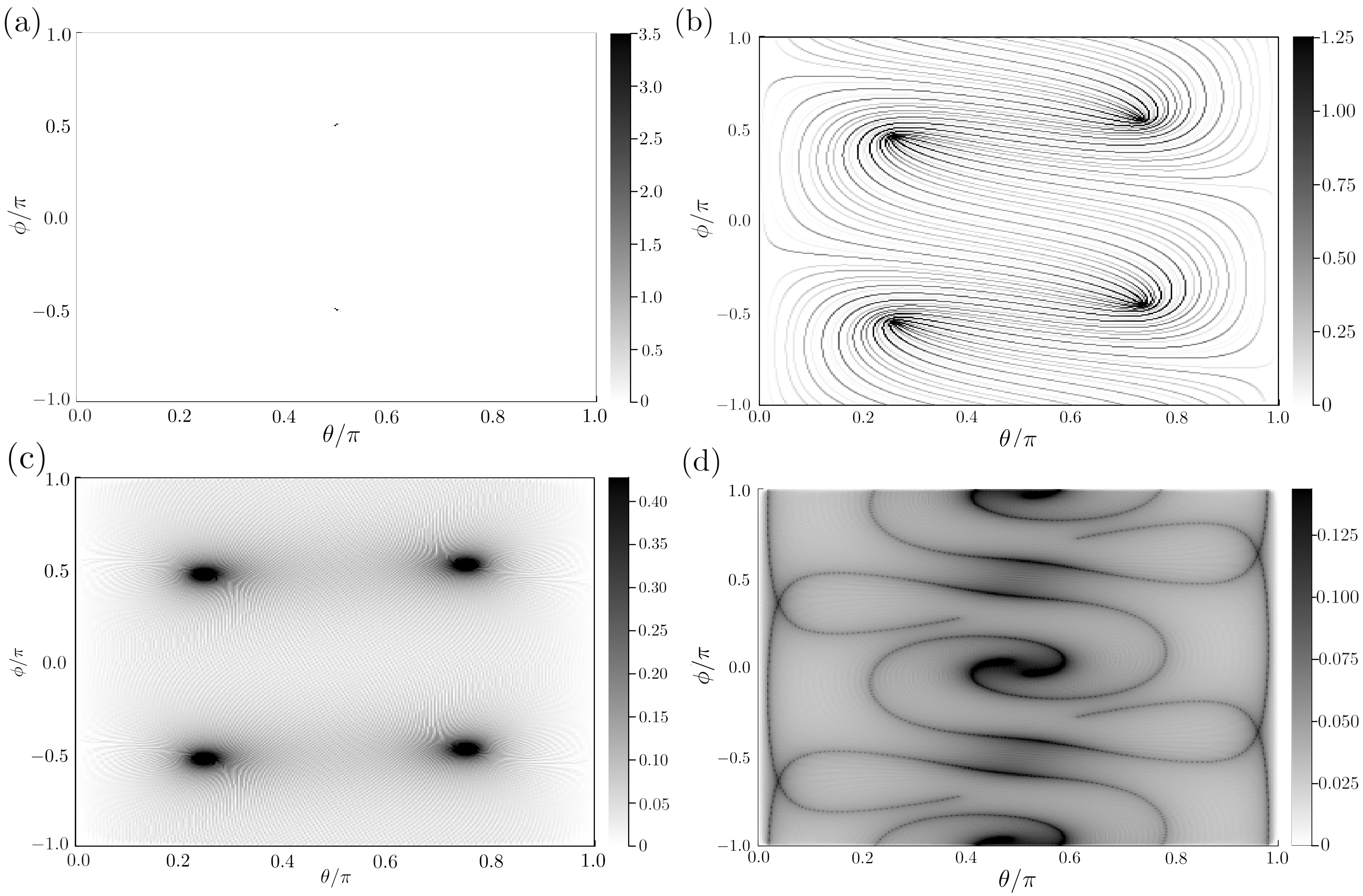}
\caption{Probability distribution functions for the output mode over angles 
$\theta$ and $\phi$ on the Bloch sphere. Parameters for the four panels are 
(a) $\beta=0.04$ and $\theta_r=\pi/4$, and yield highly localized Pauli-Y
eigenstates; (b) $\beta=0.01$ and $\theta_r=0.0681\pi$; (c) $\beta=0.001$ and 
$\theta_r=0.38467\pi$ yielding non-Pauli eigenstates with high probability; and 
(d) $\beta=0.001$ and applied rotation angles in the range 
$\theta_r\in[0.38012\pi, 0.38248\pi]$ so that the four symmetry-connected 
points on the Bloch sphere form trajectories.}
\label{fig:dancing_ghosts_map}
\end{figure*}

\noindent \textit{Eigenstates at Finite Damping}---
For small but non-zero damping, $\beta\ll 1$, one can expand $z$ in a Taylor 
series to lowest order in $\beta$; the Gaussians appearing in 
Eq.~(\ref{eq:theta_rational}) then become
\begin{equation}
\exp\left[-\frac{(z + \ell \pi / 2v)^2}{\pi \beta'}\right]\to
e^{-(k-k_m)^2\pi/2\beta}e^{-i\pi k(k-k_m)u/v}.
\end{equation}
The second exponential represents a phase proportional to the distance between 
the measured value of rescale position quadrature $k$ and the location $k_m$
of the Gaussian peak; this is much 
smaller than the first exponential and can be ignored. For any fixed choice of 
either $u/v$ or $v/u$, the Gaussians are highly peaked at locations
$k_m=\left(\ell/\sqrt{(u/v)^2+1}\right)/v$ or $k_m=\left(\ell/\sqrt{(v/u)^2+1}
\right)/u$. These are well-separated for sufficiently small $\beta$ and value 
of $u$ or $v$, and the output will again correspond to Pauli eigenstates. As 
$\beta$ increases at constant $v$ ($u$), or $v$ ($u$) increases at constant 
$\beta$, the spacing between successive peaks $k_{m+1}-k_m$ decreases. The sum
over $\ell$ in Eq.~(\ref{eq:theta_rational}) will no longer be dominated by a
single peak, and the output state will begin to deviate from a Pauli 
eigenstate. Thus, obtaining Pauli eigenstates requires a tradeoff between the
peak width (governed by $\beta$) and peak separation (governed by the magnitude
of $u$ or $v$): a coarse rational representation of the applied phase
$\theta_r$, with $u,v\sim 1$, will almost certainly yield Pauli eigenstates for 
some small value of $\beta$ and vice versa; whereas Pauli-state outcomes are 
likely to result for a finer representation of $\theta_r$, with $u,v\gg 1$, 
only for much smaller values for $\beta$.

Given the complexity introduced by the Jacobi theta functions and the
non-invertible nature of the Bloch parameters $\theta(q_m)$ and
$\phi(q_m)$, the final probability distributions are determined
numerically without approximation (refer to the Appendix for additional 
details).
For a fixed rotation angle $\theta_r$, the two-mode state after Fock
damping and the rotation on mode~1, and before measurement, is denoted
$\ket{\Psi_2}$.  A $q$-homodyne measurement on mode~1 with outcome $q_m$
then occurs with probability density
\begin{equation}
  P(q_m)
  = \frac{\big\|(\bra{q_m}\otimes \mathbb{1})\ket{\Psi_2}\big\|^2}
         {\langle\Psi_2|\Psi_2\rangle}.
  \label{eq:P_compact}
\end{equation}
The induced distributions on the Bloch sphere are obtained by
grid-based numerical pushforward method~\cite{Evans2005}.  For each
value of $\beta$ the symmetric homodyne window
$[q_{\min},q_{\max}]$ is adapted so that the cumulative probability
inside the window satisfies
$ \int_{|q_m|<q_{\max}} P(q_m)\, \mathrm{d}q_m
\;\gtrsim\; 1 - 10^{-7}$;
this guarantees that the tail probability outside the window is
$P_{\rm tail}\lesssim 10^{-7}$.  The interval $[q_{\min},q_{\max}]$ is
then discretized on a dense uniform grid with spacing
$\Delta q = (\sqrt{\pi})/\text{q\_dens}$, with typically
$\text{q\_dens}=10^5$ points per unit $\sqrt{\pi}$.  At each grid point the corresponding Bloch parameters are obtained numerically via
\begin{eqnarray}
  \phi &=& \tan^{-1}\!\left(\frac{{\rm Im}\,C_-}{{\rm Re}\,C_-}\right)
           -\tan^{-1}\!\left(\frac{{\rm Im}\,C_+}{{\rm Re}\,C_+}\right),
           \nonumber\\
  \theta &=& 2\tan^{-1}\!\left(\frac{|C_-|}{|C_+|}\right),
  \label{eq:Blochangles}
\end{eqnarray}
and a probability weight $P(q_{m,j})\,\Delta q$ is assigned to that
$(\theta,\phi)$ value.  Aggregating these weights over the grid yields
the marginal probability density functions $g(\theta)$ and $g(\phi)$, and
the two-dimensional Bloch-sphere PDFs are plotted in 
Fig.~\ref{fig:dancing_ghosts_map}. Details of the adaptive-window criterion and 
grid-resolution convergence are given in the Appendix.
 
The theory presented above indicates that Pauli eigenstates will be 
overwhelmingly likely for some small but finite $\beta$ and the smallest 
possible choices for the elements comprising the rational fraction, 
$u,v\in\{0,1\}$. These correspond to rotation angles $\theta_r=n\pi/4$, 
$n=0,\ldots,4$. Indeed, the PDF for $\theta_r=\pi/4$ and $\beta=0.04$ shown in 
Fig.~\ref{fig:dancing_ghosts_map}(a) consists of two sharp peaks on the Bloch
sphere at $\theta=\pi/2$ and $\phi=\pm\pi/2$, which match to the two orthogonal 
Pauli-Y eigenstates, consistent with the predictions based on the Jacobi theta 
function analysis ($u=v=1$, both odd). These peaks further sharpen for smaller 
$\beta$, and quickly become indistinguishable from a single point; even though
measurement outcomes obtained away from the Gaussian maxima are not 
insignificant, they yield the same output states. For larger values of $\beta$, 
the PDF peaks become more smeared out, primarily in the $\theta$ direction. 
Similar results are obtained for other Pauli eigenstate outputs. Notably, this 
protocol permits the teleportation of all Clifford gates on Fock-damped sensor
states with near unit probability for any value of $\beta\lesssim 0.04$, with 
no additional loss in the qubit quality.

To compare with experimental squeezing levels we convert $\beta$ to an
\emph{effective per-peak squeezing in dB}.  As discussed in
Refs.~\cite{Duivenvoorden_2017,Hastrup_2023}, the non-unitary map
$\rho\mapsto e^{-\beta\hat n}\rho\,e^{-\beta\hat n}$ corresponds, in the
Wigner representation, to convolving the state with an isotropic Gaussian
kernel, which replaces each Dirac peak of the ideal GKP comb by a
Gaussian peak with quadrature variance
\begin{equation}
  \sigma^2(\beta) = \tfrac{1}{2}\tanh\beta.
  \label{eq:peak_variance}
\end{equation}
Following the usual convention, we quote the associated per-peak
squeezing relative to the vacuum variance
$\sigma_{\rm vac}^2=\tfrac12$ as
\begin{equation}
  \Delta_{\mathrm{dB}}
  = -10\log_{10}\!\bigl(2\sigma^2(\beta)\bigr)
  = -10\log_{10}\!\bigl(\tanh\beta\bigr),
  \label{eq:beta_to_db}
\end{equation}
so that $\Delta_{\mathrm{dB}}>0$ corresponds to peaks that are narrower
(more squeezed) than vacuum. Thus, $\beta \approx$  0.04 corresponds to approximately 14 dB of per-peak squeezing.

Of particular interest is the possibility of obtaining non-Pauli eigenstates
of damped GKP states following this protocol. Consider the PDF shown in 
Fig.~\ref{fig:dancing_ghosts_map}(b), with parameters $\beta=0.01$ (20~dB of
per-peak squeezing) and $\theta_r=0.0681\pi$. The results clearly demonstrate 
that possible output states can be found throughout the Bloch sphere, 
manifested by long `strings' of probability, a behavior that is found for all 
choices of $\theta_r$ derived from values of $u,v\gg 1$ for this value of 
$\beta$. For these specific parameters, significant concentration on the Bloch 
sphere is centered at $(\theta,\phi)\approx(\pi/4,\phi)$, close to standard 
magic states defined by $\theta=\pi/4,3\pi/4$ and $\phi=\pm\pi/2$. 
The patterns on the Bloch sphere are located symmetrically under the following 
transformations (more details are provided in the Appendix): 
$(\theta,\phi)\rightarrow
(\theta,\phi-\pi)$, $(\pi-\theta,-\phi)$, and $(\pi-\theta,\pi-\phi)$, which 
correspond to logical $Z$, $X$, and $Y$, respectively; given that the output 
location on the Bloch sphere is heralded from the measurement outcome through 
Eqs.~(\ref{eq:Cplusminus}) and (\ref{eq:Blochangles}), these correspond to the 
CV analogs of Pauli byproduct operators in MBQC. The numerical results 
therefore vindicate the prediction from the analytics that non-Pauli 
eigenstates generically result from this protocol at finite Fock damping, when 
$u,v\gg 1$.

Although the process is inherently probabilistic, the results remain remarkably favorable even when the Fock damping is comparatively larger ($
\beta=0.01$). As shown in Fig.~\ref{fig:dancing_ghosts_map}(b), for these parameters, measurement of the first mode yields an output state with fidelity $F \gtrsim 0.94$ to one of the four Pauli-related magic states in nearly $50\%$ of the cases. If we instead require a more stringent fidelity threshold of $F \gtrsim 0.999$, the success probability decreases to about $2\%$. Reducing the damping enhances this performance: for smaller $\beta$, the probability of obtaining a high-fidelity magic state ($F \gtrsim 0.999$) increases to approximately $6\%$, and the output distribution becomes more tightly localized around the target, as seen in Fig.~\ref{fig:dancing_ghosts_map}(c) for $\beta=0.001$ (corresponding to 30~dB of per-peak squeezing) and $\theta_r = 0.38467\pi$. These results highlight that, while the generation is not fully deterministic, tuning $\theta_r$ and the rational approximation $u/v$ enables targeted non-Pauli outputs with high probability. Though 30~dB of squeezing exceeds current experimental capabilities, the trend clearly illustrates the strong benefits of advancing toward this regime. 
To highlight the practicality of this approach, even within currently feasible experimental conditions --- corresponding to achievable squeezing levels of about 14~dB ($\beta = 0.04$) --- approximately $40\%$ of $q$-homodyne measurement outcomes yield output states with fidelity $F \gtrsim 0.96$ to one of the four Pauli-related magic states. This fidelity is well above the thresholds required for magic-state distillation protocols~\cite{Bravyi2005,Meier2013,Jones2013}, indicating that the generated states are directly useful as raw resources for fault-tolerant non-Clifford gate implementation. 

With sufficiently small $\beta$, almost any non-Pauli output state can be 
targeted with high probability by a suitable choice of $\theta_r$. Consider, 
for example, $\beta=0.001$ and $\theta_r\in[0.38012\pi,0.38248\pi]$, an 
interval that includes no Pauli output states. As the value of $\theta_r$ is 
varied, the localized peaks in the PDF change their positions continuously on 
the Bloch sphere, leaving a trajectory that is depicted in 
Fig.~\ref{fig:dancing_ghosts_map}(d). Completely different trajectories are
found by choosing different intervals. In the absence of a closed-form 
expression, a numerical search is required in practice to obtain the best 
$\theta_r$ value for a given $\beta$ that yields the desired output parameters 
$(\theta,\phi)$.
Although an analytic prediction of the final Bloch-sphere location remains elusive, numerical evaluation of the governing expressions identifies the attainable points for specific parameter sets, each connected to its symmetry-related counterparts through Pauli operations.

\noindent \textit{Discussion and Outlook}--- 
The analysis shows that the finite-energy envelope of realistic GKP
states, modeled here via Fock damping with parameter \(\beta>0\), does not 
merely approximate the ideal code, or 
constitute a source of noise. Rather, it supplies a key ingredient for 
computational universality within the framework of Gaussian operations and 
homodyne measurements: small but finite damping permits the teleportation of
both Clifford and non-Pauli gates with high probability. These results provide
a springboard to a range of promising research directions. These include:
how to optimize the parameters to yield non-Pauli gates with higher fidelity;
how to prepare specific non-Pauli states with high probability and fidelity at
experimentally accessible squeezing levels (i.e., higher values of $\beta$);
and how to extend the circuit to multiple modes in order to perform more complex gates or to optimize the parameters. These and related questions will be pursued in future
work.

\noindent \textit{Acknowledgments}---The authors acknowledge Mear M. R. Koochakie for his helpful discussions and numerical assistance.

\newpage

\begin{widetext}
\appendix

\begin{center}
{\bf Appendix}
\end{center}

\section{Derivation of the Output State}

\mmn The initial state is taken to be
\begin{equation}
|\Psi_1\rangle=CZ|+\rangle|+\rangle=\frac{1}{\sqrt{2}}\left(|0\rangle|+\rangle
+|1\rangle|-\rangle\right),
\end{equation}
where $|\pm\rangle$ are ideal (undamped) sensor states (squeezed $|+\rangle$
GKP states). This can be re-expressed as
\begin{eqnarray}
|\Psi_1\rangle&=&\frac{1}{\sqrt{2}}\left(\sum_{j=-\infty}^{\infty}|2j\sqrt{\pi}
\rangle_q|+\rangle+\sum_{j=-\infty}^{\infty}|(2j+1)\sqrt{\pi}\rangle_q
|-\rangle\right)\nonumber \\
&=&\frac{1}{\sqrt{2}}\left(\sum_{j=-\infty}^{\infty}\sum_{n=0}^{\infty}
\psi_n(2j\sqrt{\pi})|n\rangle|+\rangle+\sum_{j=-\infty}^{\infty}
\sum_{n=0}^{\infty}\psi_n[(2j+1)\sqrt{\pi}]|n\rangle|-\rangle\right),
\end{eqnarray}
where the harmonic oscillator wavefunctions are given by
\begin{equation}
\psi_n(x)=\tfrac{1}{\sqrt{2^n n!}}\pi^{-1/4}e^{-x^2/2}H_n(x)
\end{equation}
and $H_n(x)$ are Hermite polynomials. Note that these states are not formally 
normalizeable, but their Fock-damped counterparts below are.

Applying the rotation to the first mode via
\begin{equation}
e^{i\theta_r\hat{n}}|q\rangle=\sum_{n=0}^{\infty}\psi_n(q)e^{i\theta_r n}
|n\rangle
\end{equation}
and Fock damping to both modes via
\begin{equation}
e^{-\beta\hat{n}}|q\rangle=\sum_{n=0}^{\infty}\psi_n(q)e^{-\beta n}|n\rangle,
\end{equation}
the state becomes
\begin{equation}
|\Psi_2\rangle=\frac{1}{\sqrt{2}}\left(\sum_{j=-\infty}^{\infty}
\sum_{n=0}^{\infty}\psi_n(2j\sqrt{\pi})e^{i\theta_r-\beta}|n\rangle|\tilde{+}
\rangle+\sum_{j=-\infty}^{\infty}\sum_{n=0}^{\infty}\psi_n[(2j+1)\sqrt{\pi}]
e^{i\theta_r-\beta}|n\rangle|\tilde{-}\rangle\right),
\end{equation}
where $|\tilde{\pm}\rangle$ represent Fock-damped sensor states. Applying a
$q$-homodyne measurement to the first mode yields
\begin{equation}
\langle q|\Psi_2\rangle=\frac{1}{\sqrt{2}}\left(\sum_{j=-\infty}^{\infty}
\sum_{n=0}^{\infty}\psi_n(2j\sqrt{\pi})e^{i\theta_r-\beta}\psi_n(q)
|\tilde{+}\rangle+\sum_{j=-\infty}^{\infty}\sum_{n=0}^{\infty}
\psi_n[(2j+1)\sqrt{\pi}]e^{i\theta_r-\beta}\psi_n(q)|\tilde{-}\rangle\right),
\end{equation}
which can be alternatively expressed as 
\begin{equation}
    \ket{\psi}_{\text{out}} = 
    \frac{1}{\mathcal{N}}
    \Big(
    \langle q_m|\tilde{0}\rangle_r \ket{\tilde{+}} +
    \langle q_m|\tilde{1}\rangle_r \ket{\tilde{-}}
    \Big),
\end{equation}
where the coefficients are given by
\begin{eqnarray}
\langle q_m|\tilde{s}\rangle_r&=&\sum_{j=-\infty}^{+\infty} \sum_{n=0}^{\infty}
\psi_n\!\big((2j+s)\sqrt{\pi}\big)\,e^{(i\theta_r - \beta)n}\,\psi_n(q_m),\;
s\in\{0,1\}\nonumber \\
&=&\frac{e^{-q_m^2/2}}{\sqrt{\pi}}\sum_{j=-\infty}^{+\infty}e^{-(2j+s)^2\pi/2}
\sum_{n=0}^{\infty}\frac{e^{(i\theta_r - \beta)n}}{2^n n!}
H_n\!\big((2j+s)\sqrt{\pi}\big) H_n(q_m).
\end{eqnarray}
The summation over $n$ may be evaluated using Mehler's formula:
\begin{equation}
\sum_{n=0}^{\infty}\frac{H_n(q)H_n(q')}{n!}\left(\frac{w}{2}\right)^n
=\frac{1}{\sqrt{1-w^2}}\exp\!\left(\frac{2qq'w-(q^2+q'^2)w^2}{1-w^2}\right),
\label{eq:mehler}
\end{equation}
which is valid for $|w|<1$. In the present case, $w = e^{i\theta_r-\beta}$, and 
the convergence condition is satisfied since $|e^{i\theta_r-\beta}|<1$ for
$\beta>0$. Note that this work is careful to always consider finite
$\beta$, and then to take the limit $\beta\to 0$. One obtains
\begin{equation}
\langle q_m|\tilde{s}\rangle_r\propto\frac{e^{-q^2/2}e^{-q^2/(w^{-2}-1)}}
{\sqrt{1-w^2}}\sum_{j=-\infty}^{+\infty}e^{-{q'}^2/2}\exp\left(\frac{2qq'w
-{q'}^2w^2}{1-w^2}\right),
\end{equation}
where $q'=2j\sqrt{\pi}$ and $(2j+1)\sqrt{\pi}$ for $s=0$ and 1, respectively,
and $w=e^{i\theta_r-\beta}$. After some straightforward algebraic 
manipulations, one obtains
\begin{equation}
\langle q_m|\tilde{s}\rangle_r\propto\frac{e^{i\frac{\pi}{2}k^2\tan\zeta}}
{2(1+e^{2i\zeta})}\vartheta_3\left\{-\frac{k\pi}{2}\sec\zeta+s\frac{\pi}{2},
e^{i\frac{\pi}{2}\tan\zeta}\right\},
\end{equation}
where $k=q_m/\sqrt{\pi}$, $\zeta\equiv\theta_r+i\beta$, and $\vartheta_3$ is
the Jacobi theta function of the third kind, defined by
\begin{equation}
\vartheta_3\{z,\omega\}=1+2\sum_{\ell=1}^{\infty}\omega^{\ell^2}\cos(2\ell z).
\label{eq:theta3}
\end{equation}
Note that $\vartheta_4\{z,\omega\}=\vartheta_3\{z+\frac{\pi}{2},\omega\}$, so
that the coefficients of the output state $|\Psi\rangle_{\rm out}
=C_+|\tilde{+}\rangle+C_-|\tilde{-}\rangle$ are (ignoring common factors)
\begin{equation}
C_+\propto\vartheta_3\left\{-\frac{k\pi}{2}\sec\zeta,
e^{i\frac{\pi}{2}\tan\zeta}\right\}:=\vartheta_3\{z,\omega\}
=\vartheta_3\{z|\tau\};\;
C_-\propto\vartheta_4\left\{-\frac{k\pi}{2}\sec\zeta,
e^{i\frac{\pi}{2}\tan\zeta}\right\}:=\vartheta_4\{z,\omega\}
=\vartheta_4\{z|\tau\},
\end{equation}
where
\begin{equation}
z=-\frac{\pi}{2}k\,\sec\zeta;\quad\omega=e^{\frac{i}{2}\pi\tan\zeta}
\equiv e^{i\pi\tau}\quad\Rightarrow\quad\tau=\frac{1}{2}\tan\zeta.
\label{eq:defs}
\end{equation}

\subsection{Symmetries of the Output State}

\mmn 
The PDF on the Bloch sphere, shown in Fig.~2
of the
main text, reveals that the output state satisfies the symmetries 
$(\theta,\phi)\rightarrow(\theta,\phi-\pi)$, $(\pi-\theta,-\phi)$, and 
$(\pi-\theta,\pi-\phi)$. These are now proven using the properties of the 
Jacobi theta functions.

The Jacobi theta functions defined by Eq.~(\ref{eq:theta3}) have two 
other convenient representations:
\begin{equation}
\vartheta_3\{z,\omega\}=1+\sum_{\ell=1}^{\infty}\omega^{\ell^2}
\left(e^{2i\ell z}+e^{-2i\ell z}\right)
=\sum_{\ell=-\infty}^{\infty}\omega^{\ell^2}e^{-2i\ell z}
=\sum_{\ell=-\infty}^{\infty}e^{i\pi\tau\ell^2}e^{-2i\ell z};
\label{eq:theta3b}
\end{equation}
and
\begin{equation}
\vartheta_3\{z,\omega\}=\frac{1}{\sqrt{-i\tau}}\sum_{\ell}
e^{-i\pi(z/\pi\pm \ell)^2/\tau}
=\frac{1}{\sqrt{-i\tau}}e^{-iz^2/\pi\tau}\sum_{\ell}
e^{-i\pi\ell^2/\tau\mp 2i\ell z/\tau}.
\label{eq:theta3c}
\end{equation}
To prove Eq.~(\ref{eq:theta3c}), rewrite Eq.~(\ref{eq:theta3b}) as follows:
\begin{eqnarray}
\vartheta_3\{z,\omega\}&=&\sum_{\ell}\int_{-\infty}^{\infty}
e^{i\pi\tau x^2}e^{-2izx}\delta(x-\ell)dx
=\int_{-\infty}^{\infty}e^{i\pi\tau x^2}e^{-2izx}\sum_{\ell}\delta(x-\ell)dx
\nonumber \\
&=&\int_{-\infty}^{\infty}
e^{i\pi\tau x^2}e^{-2izx}\sum_{r} e^{\pm 2i\pi rx}dx
=\sum_{\ell}\int_{-\infty}^{\infty}e^{i\pi\tau x^2}e^{-2izx}
e^{\pm 2i\pi\ell x}dx\nonumber \\
&=&\sum_{\ell}\int_{-\infty}^{\infty}e^{i\pi\tau\left[x^2-2x(z/\pi\pm \ell)
/\tau\right]}dx.
\label{eq:neattrick}
\end{eqnarray}
Complete the square, and perform the Gaussian integral:
\begin{eqnarray}
\vartheta_3\{z,\omega\}&=&\sum_{\ell}e^{-i\pi\tau(z/\pi\pm \ell)^2/\tau^2}
\int_{-\infty}^{\infty}
e^{i\pi\tau\left[x-(z/\pi\pm \ell)/\tau\right]^2}dx
=\sum_{\ell}e^{-i\pi\tau(z/\pi\pm \ell)^2/\tau^2}\frac{1}{\sqrt{-i\tau}}
\nonumber \\
&=&\frac{1}{\sqrt{-i\tau}}e^{-iz^2/(\pi\tau)}
\sum_{\ell}e^{-i\pi\tau(\pm 2\ell z/\pi+\ell)^2/\tau^2}
=\frac{1}{\sqrt{-i\tau}}e^{-iz^2/\pi\tau}\sum_{\ell}
e^{-i\pi\ell^2/\tau\mp 2i\ell z/\tau}.
\nonumber
\end{eqnarray}
which is valid as long as Im$(\tau)>0$. In our case for $\beta\ll 1$
$$\tau=\frac{1}{2}\tan\zeta\approx\frac{1}{2}\left[\tan\theta_r
+i\beta\left(1+\tan^2\theta_r\right)\right],$$
and so the condition is satisfied because $\beta>0$. Finally, the 
Jacobi theta function is invariant on $z\to -z$. This completes the proof.

The Jacobi elliptic functions have three convenient symmetries. These are
\begin{eqnarray}
\vartheta_3\left\{z+(m+n\tau)\pi,\omega\right\}
&=&q^{-n^2}e^{-2inz}\vartheta_3\{z,\omega\},\quad m,n\in\mathbb{Z};
\label{eq:sym1} \\
\vartheta_3\left\{z\pm\frac{\pi}{2}\right|\left.\vphantom{\frac{\pi}{2}}\tau
\right\}&=&\vartheta_3\{z|\tau\pm 1\}\quad\Leftrightarrow\quad
\vartheta_3\left\{z\pm\frac{\pi}{2}\right|\left.\vphantom{\frac{\pi}{2}}\tau
\pm 1\right\}=\vartheta_3\left\{z|\tau\right\};
\label{eq:sym2} \\
\vartheta_3\left\{\frac{z}{\tau}\right|\left. -\frac{1}{\tau}\right\}
&=&e^{iz^2/(\pi\tau)}\sqrt{-i\tau}\vartheta_3\left\{z|\tau\right\}.
\label{eq:sym3}
\end{eqnarray}
Consider first symmetry~(\ref{eq:sym1}). The case with $m\neq 0$ and $n=0$ 
follows directly from Eq.~(\ref{eq:theta3}):
$$\vartheta_3\{z+m\pi,\omega\}=1+2\sum_{\ell=1}^{\infty}\omega^{\ell^2}
\cos(2\ell z+2\ell m\pi)=\vartheta_3\{z,\omega\}.$$
The case with $m=0$ and $n\neq 0$ is not as obvious. Using 
Eq.~(\ref{eq:theta3c}), the transformation $z\to z+n\pi\tau$ yields
\begin{eqnarray}
\vartheta_3\left\{z+n\pi\tau|\tau\right\}&=&\sqrt{\frac{\tau}{i}}
e^{-i(z+n\pi\tau)^2/\pi\tau}\sum_{j=-\infty}^{\infty}e^{-i\pi j^2/\tau}
e^{-2ij(z+n\pi\tau)/\tau}\nonumber \\
&=&\sqrt{\frac{\tau}{i}}e^{-iz^2/\pi\tau}e^{-in^2\pi\tau}
e^{-2inz}\sum_{j=-\infty}^{\infty}e^{-i\pi j^2/\tau}
e^{-2ijz/\tau}
=\omega^{-n^2}e^{-2inz}\vartheta_3\left\{z|\tau\right\},
\end{eqnarray}
as needed.

Symmetry~(\ref{eq:sym2}) again follows directly from the definition of the 
theta functions, Eq.~(\ref{eq:theta3}):
\begin{eqnarray}
\vartheta_3\left\{z\pm\frac{\pi}{2}\right|\left.\vphantom{\frac{\pi}{2}}\tau
\right\}&=&1+2\sum_{\ell=1}^{\infty}\omega^{\ell^2}\cos\left(2\ell z+\ell\pi
\right)=1+2\sum_{\ell=1}^{\infty}(-1)^{\ell}\omega^{\ell^2}\cos(2\ell z)
=1+2\sum_{\ell=1}^{\infty}(-\omega)^{\ell^2}\cos(2\ell z)
\nonumber \\
&=&1+2\sum_{\ell=1}^{\infty}\left(e^{i\pi(\tau\pm 1)}\right)^{\ell^2}
\cos(2\ell z)=\vartheta_3\{z|\tau\pm 1\}.
\label{eq:sym2proof}
\end{eqnarray}
Note that this also implies that $\vartheta_3\{z\pm\frac{\pi}{2},\omega\}
=\vartheta_3\{z,-\omega\}$. Combining this with symmetry~(\ref{eq:sym1}) gives
$$\vartheta_3\left\{z+\frac{\pi}{2}+\frac{\pi}{2}\right|\left.
\vphantom{\frac{\pi}{2}}\tau\right\}=\vartheta_3\{z|\tau\}
=\vartheta_3\left\{z+\frac{\pi}{2}\right|\left.
\vphantom{\frac{\pi}{2}}\tau\pm 1\right\}=\vartheta_3\{z|\tau\pm 2\},$$
which implies that $\vartheta_3\{z|\tau\}$ is periodic on $\tau\to\tau\pm 2$.
The same conclusion could also have been reached directly from the fact that
$\omega=e^{i\pi\tau}$ so that $\tau\to\tau\pm 2$ maps $\omega\to\omega$. 

Finally, consider symmetry~(\ref{eq:sym3}). Using Eq.~(\ref{eq:theta3b})
one obtains
$$\vartheta_3\left\{\frac{z}{\tau}\right|\left. -\frac{1}{\tau}\right\}
=\sum_{\ell}e^{-i\pi\ell^2/\tau}e^{-2i\ell z/\tau}.$$
Comparing the right-hand side with Eq.~(\ref{eq:theta3c}) immediately yields
$$\vartheta_3\left\{\frac{z}{\tau}\right|\left. -\frac{1}{\tau}\right\}
=\sqrt{-i\tau}e^{iz^2/\pi\tau}\vartheta_3\{z|\tau\},$$
which completes the proof.

If $z\to z+\pi/2$, then $\vartheta_3\{z|\tau\}\to
\vartheta_3\{z+\frac{\pi}{2}|\tau\}$, corresponding to $C_+\to C_-$, and 
$$\vartheta_3\left\{z+\frac{\pi}{2}\right|\left.\vphantom{\frac{\pi}{2}}\tau
\right\}\to\vartheta_3\{z+\pi|\tau\}=\vartheta_3\{z|\tau+2\}
=\vartheta_3\{z|\tau\},$$
using symmetry~(\ref{eq:sym2}), corresponding to $C_-\to C_+$. Under this 
transformation, the new angles on the Bloch sphere become
\begin{eqnarray}
\theta'&=&2\tan^{-1}\left(\frac{|C_+|}{|C_-|}\right)=2\left[\frac{\pi}{2}-
\tan^{-1}\left(\frac{|C_-|}{|C_+|}\right)\right]=\pi-\theta;\nonumber \\
\phi'&=&\tan^{-1}\left(\frac{{\rm Im}(C_+)}{{\rm Re}(C_+)}\right)
-\tan^{-1}\left(\frac{{\rm Im}(C_-)}{{\rm Re}(C_-)}\right)=-\phi,\nonumber
\end{eqnarray}
which accounts for the mapping $(\theta,\phi)\to(\pi-\theta,-\phi)$.
Because $\vartheta_3\{z\pm m\pi|\tau\}=\vartheta_3\{z|\tau\}$, it is also true 
that the map $\vartheta_3\{z|\tau\}\to\vartheta_3\{z+\frac{\pi}{2}|\tau\}$ is 
equivalent to $\vartheta_3\{z+1|\tau\}\to\vartheta_3\{z+\frac{\pi}{2}|\tau\}$, 
which is the same as $\vartheta_3\{z|\tau\}\to\vartheta_3\{z-\frac{\pi}{2}
|\tau\}$. Therefore the map $C_+\leftrightarrow C_-$ corresponds more generally 
to $z\to z\pm\frac{\pi}{2}$.

Using the definition of $z$, Eq.~(\ref{eq:defs}), the map
$z\leftrightarrow z\pm\pi/2$ corresponds to
$$-\frac{\pi}{2}k\sec\zeta\to -\frac{\pi}{2}\left(k\sec\zeta\mp 1\right),$$
which is equivalent to $k\to k\pm\cos\zeta$. In other words, if an outcome $k$
is obtained for some $\theta_r$ and $\beta$ with high probability, then we 
would expect equally likely outcomes for $k\pm\cos\zeta$, or in original $q$
units $q\pm\sqrt{\pi}\cos\zeta$. Perhaps even more important, the output state 
is guaranteed to repeat on the interval $q\to q\pm 2\sqrt{\pi}\cos\zeta
\approx q\pm 2\sqrt{\pi}\cos\theta_r$ for any $\theta_r\gg\beta$. This 
observation is consistent with the high-symmetry point $\theta_r=0$, where the
output is $|+\rangle$, characterized by peaks in $q$ space at intervals of 
$2n\sqrt{\pi}$ from the $|0\rangle$ state and $(2n+1)\sqrt{\pi}$ from the 
$|1\rangle$ state.

The only transformations that can change the relative phase of $C_+$ and 
$C_-$ are the symmetry~(\ref{eq:sym1}) for $n\neq 0$ and 
symmetry~(\ref{eq:sym3}). Let's first consider symmetry~(\ref{eq:sym1}), 
corresponding to the transformation
\begin{equation}
z\to z\pm n\pi\tau\quad\Leftrightarrow\quad -\frac{\pi}{2}k\sec\zeta\to
-\frac{\pi}{2}k\sec\zeta\pm\frac{n\pi}{2}\tan\zeta\to -\frac{\pi}{2}\sec\zeta
\left(k\pm n\sin\zeta\right),
\label{eq:ztransform}
\end{equation}
or $k\to k\pm n\sin\zeta$. Then
\begin{eqnarray}
C_+[k,\theta_r,\beta]=\vartheta_3\{z,\omega\}
&\to& \vartheta_3\{z+n\pi\tau,\omega\}
=\omega^{-n^2}e^{-2inz}\vartheta_3\{z,\omega\}=e^{-i\frac{\pi}{2}n^2\tan\zeta}
e^{in\pi k\sec\zeta}C_+[k,\theta_r,\beta];\nonumber \\
C_-[k,\theta_r,\beta]=\vartheta_3\left\{z+\frac{\pi}{2},\omega\right\}&\to&
\vartheta_3\left\{z+\frac{\pi}{2}+n\pi\tau,\omega\right\}
=\omega^{-n^2}e^{-2in(z+\pi/2)}\vartheta_3\left\{z+\frac{\pi}{2},\omega\right\}
\nonumber \\
&=&(-1)^ne^{-i\frac{\pi}{2}n^2\tan\zeta}e^{in\pi k\sec\zeta}
C_-[k,\theta_r,\beta].\nonumber
\end{eqnarray}
The new prefactors are the same for both $C_+$ and $C_-$, save for an 
additional $(-1)^n$ factor on $C_-$. For even $n$ this maps 
$|\psi\rangle\to|\psi\rangle$, which implies that the measurement outcomes
are invariant under $k\pm 2n\sin\zeta$ or $q\pm 2n\sqrt{\pi}\sin\theta_r$ for
$\theta_r\gg\beta$. For small $\theta$, this is a much shorter period in $q$
than was obtained from symmetry~(\ref{eq:sym2}). For odd $n$, the 
transformation changes the sign of both the real and imaginary parts of $C_-$, 
which has no effect on $\theta$ but maps $\phi\to\phi-\pi$ using the elementary
properties of the tangent function. This recovers the map $(\theta,\phi)
\to(\theta,\phi-\pi)$. Combining the results above with this one yields the 
third and final map $(\theta,\phi)\to(\pi-\theta,\pi-\phi)$.

In summary, if the output probability is high for outcome $q$, we 
generically expect high-probability outputs at points 
\begin{eqnarray}
k&\rightarrow& k\pm\left(m\cos\zeta+n\sin\zeta\right)\approx k\pm\left(
m\cos\theta_r+n\sin\theta_r\right),\quad n,m\in\mathbb{Z};\nonumber \\
q&\rightarrow& q\pm\sqrt{\pi}\left(m\cos\zeta+n\sin\zeta\right)
\approx q\pm\sqrt{\pi}\left(m\cos\theta_r+n\sin\theta_r\right),\quad 
n,m\in\mathbb{Z},
\label{eq:series}
\end{eqnarray}
assuming that $\theta_r\gg\beta$.

\subsection{Case where $\tan\theta_r$ is rational}

\mmn Using Eqs.~(\ref{eq:defs}) and (\ref{eq:theta3b}), one can write
\begin{eqnarray}
\vartheta_3\{z,\omega\}&=&\sum_{\ell}e^{i\pi\tau\ell^2}e^{-2i\ell z}
=\sum_{\ell}e^{i\frac{\pi}{2}\ell^2\tan\zeta}e^{-2i\ell z}
=\sum_{\ell}e^{i\pi\left(\tan\theta_r
+i\beta\sec^2\theta_r\right)\ell^2/2}e^{-2i\ell z}\nonumber \\
&=&\sum_{\ell}e^{-\pi\beta\sec^2\theta_r\ell^2/2}
e^{i\pi\tan\theta_r\ell^2/2}e^{-2i\ell z}
\nonumber
\end{eqnarray}
Approximate $\tan\theta_r=u/v\mathbb{Q}$, where $u$ and $v$ are coprime 
integers. In that case,
$$\sin\theta_r=\frac{u}{\sqrt{u^2+v^2}};\;
\cos\theta_r=\frac{v}{\sqrt{u^2+v^2}}.$$
Then
$$\vartheta_3\{z,\omega\}=\sum_{\ell}e^{-\pi\beta\left(1+u^2/v^2\right)\ell^2/2}
e^{i\pi\ell^2u/2v}e^{-2i\ell z}
\equiv\sum_{\ell}e^{-\pi\beta'\ell^2}e^{i\pi\ell^2u/2v}e^{-2i\ell z},$$
where $\beta':=\left(1+u^2/v^2\right)\beta/2$ is defined for convenience.

Let's decompose the sum over $\ell=2vm+n$, where $m,n\in\mathbb{Z}$ and 
$n=0,1,\ldots,2v-1$. Consider for example $u=1$ and $v=1$; then 
$\ell=\{2m,2m+1\}$ for $n=\{0,1\}$, corresponding to summing over all 
even and odd integers. Breaking the sum up in this way is therefore equivalent
to summing over $\ell$ modulo $2v$. Then
\begin{eqnarray}
\vartheta_3\{z,q\}&=&\sum_m\sum_{n=0}^{2v-1}e^{-\pi\beta'(2vm+n)^2}
\exp\left[i\pi\left(\frac{(2vm+n)^2u}{2v}\right)\right]e^{-2i(2vm+n)z}
\nonumber \\
&=&\sum_m\sum_{n=0}^{2v-1}e^{-\pi\beta'(2vm+n)^2}
\exp\left[i\pi\left(2mu(vm+n)+\frac{n^2u}{2v}\right)\right]e^{-2i(2vm+n)z}
\nonumber \\
&=&\sum_{n=0}^{2v-1}e^{i\pi\frac{n^2u}{2v}}e^{-2inz}
\sum_m e^{-\pi\beta'(2vm+n)^2}e^{2i\pi m^2uv}e^{2i\pi mun}e^{-4ivmz}
\nonumber \\
&=&\sum_{n=0}^{2v-1}e^{i\pi\frac{n^2u}{2v}}e^{-2inz}
\sum_m e^{-\pi\beta'(2vm+n)^2}e^{-4ivmz},
\label{eq:maybe}
\end{eqnarray}
where the $e^{2i\pi m^2uv}$ and $e^{2i\pi mun}$ terms can be neglected from the 
sum over $m$, because they are both unity for all values of $m$. The sum over
$m$ can be readily evaluated:
\begin{eqnarray}
\vartheta_3\{z,q\}&=&\sum_{n=0}^{2v-1}e^{i\pi\frac{n^2u}{2v}}e^{-2inz}
\frac{1}{2v\sqrt{\beta'}}e^{-z^2/\pi\beta'}
e^{2inz}\vartheta_3\left\{\frac{n\pi}{2v}
+\frac{iz}{2v\beta'},e^{-\pi/4v^2\beta'}\right\}\nonumber \\
&=&\frac{1}{2v\sqrt{\beta'}}e^{-z^2/\pi\beta'}\sum_{n=0}^{2v-1}
e^{i\pi\frac{n^2u}{2v}}\vartheta_3\left\{\frac{n\pi}{2v}
+\frac{iz}{2v\beta'},e^{-\pi/4v^2\beta'}\right\}.
\label{eq:theta3rational}
\end{eqnarray}
Thus, a Jacobi theta function can be considered as a finite sum over other 
Jacobi theta functions. These have the parameters 
$$\tilde{z}=\frac{iz}{2v\beta'}+\frac{n\pi}{2v};\quad
\tilde{\tau}=\frac{i}{4v^2\beta'},$$
so that 
$$\frac{\tilde{z}}{\tilde{\tau}}=2v(z-i\pi n\beta');\quad
-\frac{1}{\tilde{\tau}}=4iv^2\beta';\quad 
e^{-i\tilde{z}^2/\pi\tilde{\tau}}=e^{(z-i\pi n\beta')^2/\pi\beta'}.$$
Then, using symmetry~(\ref{eq:sym3}), one obtains the equivalent expression
\begin{eqnarray}
\vartheta_3\{z,\omega\}&=&\frac{1}{2v\sqrt{\beta'}}e^{-z^2/\pi\beta'}
\sum_{n=0}^{2v-1}e^{i\pi\frac{n^2u}{2v}}2v\sqrt{\beta'}
e^{(z-i\pi n\beta')^2/\pi\beta'}
\vartheta_3\left\{2v(z-i\pi n\beta'),e^{-4\pi v^2\beta'}\right\}\nonumber \\
&=&\sum_{n=0}^{2v-1}
e^{i\pi\frac{n^2u}{2v}}e^{-2inz}e^{-\pi n^2\beta'}
\vartheta_3\left\{2v(z-i\pi n\beta'),e^{-4\pi v^2\beta'}\right\}.
\label{eq:theta3rationalb}
\end{eqnarray}
And:
\begin{eqnarray}
\vartheta_4\{z,\omega\}\equiv\vartheta_3\left\{z+\frac{\pi}{2},\omega\right\}
&=&\sum_{n=0}^{2v-1}e^{i\pi\frac{n^2u}{2v}}e^{-2inz}e^{-2in\pi/2}
e^{-\pi n^2\beta'}\vartheta_3\left\{2v\left(z+\frac{\pi}{2}-i\pi n\beta'\right),
e^{-4\pi v^2\beta'}\right\}\nonumber \\
&=&\sum_{n=0}^{2v-1}(-1)^ne^{i\pi\frac{n^2u}{2v}}e^{-2inz}e^{-\pi n^2\beta'}
\vartheta_3\left\{2v\left(z-i\pi n\beta'\right),
e^{-4\pi v^2\beta'}\right\}
\label{eq:theta3rationalc}
\end{eqnarray}
using symmetry~(\ref{eq:sym1}), because $v\in\mathbb{Z}$.

Note that in expression~(\ref{eq:theta3rationalb}), 
$\tau_{\rm eff}=4i\beta' v^2$ and $z_{\rm eff}=2vz$ for $n=0$; using 
symmetry~(\ref{eq:sym1}), one obtains
$$\vartheta_3\left\{2vz-4i\beta' v^2,e^{-4\pi v^2\beta'}\right\}
=e^{4\pi\beta' v^2}e^{4ivz}\vartheta_3\left\{2vz,e^{-4\pi v^2\beta'}\right\},$$
so that the $n=2v$ term in Eq.~(\ref{eq:theta3rationalb}) is
$$e^{i\pi\frac{4v^2u}{2v}}e^{-4ivz}e^{-4\pi v^2\beta'} 
\vartheta_3\left\{2vz-4i\pi v^2\beta',e^{-4\pi v^2\beta'}\right\}
=e^{2i\pi vu}\vartheta_3\left\{2vz,e^{-4\pi v^2\beta'}\right\}
=\vartheta_3\left\{2vz,e^{-4\pi v^2\beta'}\right\},$$
which coincides with the $n=0$ term. The cycle therefore repeats; likewise, the 
periodicity applies to Eq.~(\ref{eq:theta3rationalc}) because the period $2v$ 
is always even.

Using Eq.~(\ref{eq:neattrick}), the Jacobi theta function appearing on the 
right-hand side of Eq.(\ref{eq:theta3rationalb}) can be written as
\begin{eqnarray}
\vartheta_3\{2v(z-i\pi n\beta'),e^{-4\pi v^2\beta'}\}
&=&\sum_{\ell=-\infty}^{\infty}\frac{1}{2v\sqrt{\beta'}}\exp\left[
-\frac{\left[\ell\pi+2v(z-i\pi n\beta')\right]^2}{4\pi v^2\beta'}\right].
\nonumber \\
&=&\frac{1}{2v\sqrt{\beta'}}e^{2inz}e^{\pi n^2\beta'}
\sum_{\ell=-\infty}^{\infty}e^{i\pi \ell n/v}\exp
\left[-\frac{\left(z+\ell\pi/2v\right)^2}{\pi\beta'}\right],
\label{eq:peaks}
\end{eqnarray}
which for $\beta'\ll 1$ corresponds to strongly localized peaks centered at
$z=\ell\pi/2v$. One therefore obtains the final expressions
\begin{equation}
\vartheta_3\{z,\omega\}=\frac{1}{2v\sqrt{\beta'}}\sum_{n=0}^{2v-1}
e^{i\pi\frac{n^2u}{2v}}\sum_{\ell=-\infty}^{\infty}e^{i\pi \ell n/v}\exp
\left[-\frac{\left(z+\ell\pi/2v\right)^2}{\pi\beta'}\right],
\label{eq:theta3rationald}
\end{equation}
corresponding to Eq.~(7) in the main text, and
\begin{equation}
\vartheta_4\{z,\omega\}=\frac{1}{2v\sqrt{\beta'}}\sum_{n=0}^{2v-1}(-1)^n
e^{i\pi\frac{n^2u}{2v}}\sum_{\ell=-\infty}^{\infty}e^{i\pi \ell n/v}\exp
\left[-\frac{\left(z+\ell\pi/2v\right)^2}{\pi\beta'}\right].
\label{eq:theta3rationale}
\end{equation}

\subsection{Zero-damping limit}

\mmn For $\beta'\to 0$, the Gaussians appearing in 
Eqs.~(\ref{eq:theta3rationald}) and (\ref{eq:theta3rationale}) correspond to 
Dirac delta functions, given the definition
$$\delta(z)=\lim_{\beta\to 0}\frac{1}{\pi\sqrt{\beta}}e^{-z^2/\pi\beta};$$
one obtains
$$\vartheta_3\{2v(z-i\pi n\beta'),e^{-4\pi v^2\beta'}\}\approx
\frac{\pi}{2v}e^{2inz}\sum_{\ell=-\infty}^{\infty}e^{i\pi \ell n/v}
\delta(z+\ell\pi/2v).$$
Eqs.~(\ref{eq:theta3rationalb}) and (\ref{eq:theta3rationalc}) can then be 
written as
\begin{eqnarray}
C_+\left[k,\tan^{-1}\left(\frac{u}{v}\right),0\right]&\approx&\sum_{n=0}^{2v-1}
e^{i\pi\frac{n^2u}{2v}}e^{-2inz}e^{-\pi n^2\beta'}\frac{\pi}{2v}e^{2inz}
e^{\pi n^2\beta'}\sum_{\ell=-\infty}^{\infty}e^{i\pi \ell n/v}
\delta(z+\ell\pi/2v)
\label{eq:C+beta0} \\
&\approx &\frac{\pi}{2v}\sum_{n=0}^{2v-1}e^{i\pi\frac{n^2u}{2v}}
\sum_{\ell=-\infty}^{\infty}e^{i\pi \ell n/v}\delta(z+\ell\pi/2v);\nonumber \\
C_-\left[k,\tan^{-1}\left(\frac{u}{v}\right),0\right]&\approx&\frac{\pi}{2v}
\sum_{n=0}^{2v-1}(-1)^ne^{i\pi\frac{n^2u}{2v}}\sum_{\ell=-\infty}^{\infty}
e^{i\pi \ell n/v} \delta(z+\ell\pi/2v).\label{eq:C-beta0}
\end{eqnarray}
Eqs.~(\ref{eq:C+beta0}) and (\ref{eq:C-beta0}) clearly show that $C_+$ and 
$C_-$ will both be non-zero only for very specific values of $z$ when 
$\beta\to 0$, i.e.\ for very specific measurement outcomes. Given that 
$z=-(\pi k/2)\sec\theta_r=-(\pi k/2)\sqrt{u^2+v^2}/v$ in this limit, the output 
state is only non-zero when 
$$-\frac{\pi k}{2v}\sqrt{u^2+v^2}=-\frac{\ell\pi}{2v}\quad\Rightarrow\quad
k=\frac{\ell}{\sqrt{u^2+v^2}},\;\ell\in\mathbb{Z}.$$
It is instructive to compare this result to Eq.~(\ref{eq:series}), where the
peaks in the coefficients were found to be located at
$$k\approx\frac{nu+mv}{\sqrt{u^2+v^2}}.$$
It turns out that, indeed, the peaks in the probability appear for all integers
$\ell$.

Eqs.~(\ref{eq:C+beta0}) 
and (\ref{eq:C-beta0}) together yield
$$\frac{C_-\left[\frac{\ell}{\sqrt{u^2+v^2}},\tan^{-1}\left(\frac{u}{v}\right),
0\right]}{C_+\left[\frac{\ell}{\sqrt{u^2+v^2}},
\tan^{-1}\left(\frac{u}{v}\right),0\right]}\approx
\frac{\sum_{n=0}^{2v-1}(-1)^ne^{i\pi\frac{n^2u}{2v}}e^{i\pi \ell n/v}}
{\sum_{n=0}^{2v-1}e^{i\pi\frac{n^2u}{2v}}e^{i\pi \ell n/v}}.$$

Let's now show that only Pauli eigenstates result when $\beta\to 0$.
Ignoring constant factors, one obtains
\begin{eqnarray}
\left|\Psi\left[\frac{\ell}{\sqrt{u^2+v^2}},\tan^{-1}\left(\frac{u}{v}\right),
0\right]\right\rangle_{\rm out}
&=&\sum_{n=0}^{2v-1}e^{i\pi\frac{n^2u}{2v}}e^{i\pi \ell n/v}|+\rangle
+\sum_{n=0}^{2v-1}(-1)^ne^{i\pi\frac{n^2u}{2v}}e^{i\pi \ell n/v}|-\rangle
\nonumber \\
&=&2\sum_{n\in{\rm even}}^{2v-2}e^{i\pi\frac{n^2u}{2v}}
e^{i\pi \ell n/v}|0\rangle
+2\sum_{n\in{\rm odd}}^{2v-1}e^{i\pi\frac{n^2u}{2v}}e^{i\pi \ell n/v}|1\rangle
\nonumber \\
&=&2\sum_{n=0}^{v-1}e^{i\pi\frac{2n^2u}{v}}
e^{2i\pi \ell n/v}|0\rangle
+2\sum_{n=0}^{v-1}e^{i\pi\frac{(2n+1)^2u}{2v}}e^{i\pi \ell(2n+1)/v}|1\rangle
\nonumber \\
&=&2\sum_{n=0}^{v-1}e^{2i\pi\frac{n^2u}{v}}
e^{2i\pi \ell n/v}|0\rangle
+2e^{i\pi(2\ell+u)/2v}\sum_{n=0}^{v-1}e^{2i\pi\frac{n^2u}{v}}
e^{2i\pi n(\ell+u)/v}|1\rangle\nonumber \\
&=&2e^{-i\pi\ell^2/2uv}\sum_{n=0}^{v-1}
e^{2i\pi\left(n+\frac{\ell}{2u}\right)^2u/v}|0\rangle
+2e^{-i\pi\ell^2/2uv}\sum_{n=0}^{v-1}
e^{2i\pi\left(n+\frac{\ell}{2u}+\frac{1}{2}\right)^2u/v}|1\rangle\nonumber \\
&:=&C_0|0\rangle+C_1|1\rangle.\nonumber
\end{eqnarray}
Both coefficients $C_0$ and $C_1$ have a close resemblance to generalized 
quadratic Gauss sums
$$G(u,\ell,v)=\sum_{n=0}^{v-1}e^{2i\pi(un^2+\ell n)/v},$$
where $u,\ell,v\in\mathbb{Z}$ as has been assumed. The general solution depends 
crucially on the characteristics of $u$, $\ell$, and $v$, i.e.\ if they are 
mutually prime, even, odd, etc.; so that general results are not generally
straightforward to obtain. But certain cases are known analytically. For 
example, if $u=0$ (no rotation) and choosing $v=1$, then $C_+=1$ and $C_-=0$
for $\ell$ even and vice versa for $\ell$ odd; this corresponds to outputs of 
$|+\rangle$ and $|-\rangle$, respectively, as expected. 

If $v$ is odd, then
$$G(u,\ell,v)=\varepsilon_c\sqrt{v}\left(\frac{u}{v}\right)e^{-2i\pi
\overline{4u}\ell^2/v},$$
where
$$\left(\frac{u}{v}\right)=\begin{cases}
0 & u=0~({\rm mod}~v)\cr
1 & x^2=u~({\rm mod}~v),~x\in\mathbb{Z},~\mbox{has a solution}\cr
-1 & x^2=u~({\rm mod}~v),~x\in\mathbb{Z},~\mbox{has no solution}\cr
\end{cases}$$
is the Jacobi symbol, $\overline{4u}$ is the modular inverse of $4u$, and 
$$\varepsilon_c=\begin{cases}
1 & c=1~({\rm mod}~4)\cr
i & c=3~({\rm mod}~4).\cr
\end{cases}$$
It is straightforward to verify that under these assumptions
$$\sum_{n=0}^{v-1}e^{2i\pi\left(n+\frac{\ell}{2u}+\frac{1}{2}\right)^2u/v}
=e^{i\pi uv/2}e^{i\pi\ell}\sum_{n=0}^{v-1}e^{2i\pi\left(n+\frac{\ell}{2u}
\right)^2u/v}.$$
In this case,
$$\frac{C_1}{C_0}=e^{i\pi uv/2}e^{i\pi\ell}.$$
If $u$ and $v$ are both odd then $e^{i\pi uv/2}=\pm i$, with the measurement
outcome $\ell$ changing the sign, in which case the output state is a $Y$
eigenstate. If $u$ is even and $v$ is odd then $e^{i\pi uv/2}=\pm 1$, again
with the measurement outcome changing the sign, in which case the output state 
is an $X$ eigenstate. 

The case where $v$ is even is much trickier, and a general solution does 
not appear to exist. Instead, we turn to numerics. If $\ell$ is an even 
multiple of $u$, then $C_1=0$ for $v=0$~mod~4, and $C_0=0$ for $v=2$~mod 4. 
When $\ell$ takes other values (including odd ones), either $C_0$ or $C_1$ is 
zero, but the pattern is not clear. In all cases, however, the output 
corresponds to an eigenstate of $Z$. Therefore, for sufficiently small Fock 
damping any rational approximation to $\tan\theta_r$ leads to a Pauli 
eigenstate.

\section{Damping parameter and per-peak squeezing} 
\label{sec:squeezing}

To compare with experimental squeezing levels, $\beta$ is converted to an
\emph{effective per-peak squeezing in dB}. The non-unitary map
$\rho\mapsto e^{-\beta\hat n}\rho\,e^{-\beta\hat n}$ corresponds, in the
Wigner representation, to convolving the state with an isotropic Gaussian
kernel, which replaces each Dirac peak of the ideal GKP comb by a
Gaussian peak with quadrature variance
\begin{equation}
  \sigma^2(\beta) = \tfrac{1}{2}\tanh\beta.
  \label{eq:peak_variance}
\end{equation}
Following the usual convention, we quote the associated per-peak
squeezing relative to the vacuum variance
$\sigma_{\rm vac}^2=\tfrac12$ as
\begin{equation}
  \Delta_{\mathrm{dB}}
  = -10\log_{10}\!\bigl(2\sigma^2(\beta)\bigr)
  = -10\log_{10}\!\bigl(\tanh\beta\bigr),
  \label{eq:beta_to_db}
\end{equation}
so that $\Delta_{\mathrm{dB}}>0$ corresponds to peaks that are narrower
(more squeezed) than vacuum. Thus, for example, $\beta \approx$  0.04 
corresponds to approximately 14 dB of per-peak squeezing.

\section{Numerical methods and convergence checks} 
\label{sec:numerics}

In this section the numerical procedures used to compute the
probability density $P(q_m)$ for the $q$-homodyne outcome, to push this
distribution forward to the Bloch-sphere coordinates $(\theta,\phi)$, and
to evaluate success probabilities and fidelities are summarized. Convergence checks are presented with respect to (i) the homodyne window size (``tail probability'') and
(ii) the grid resolution in $q_m$.

\subsection{Adaptive homodyne window and tail probability}

Given the complexity introduced by the Jacobi theta functions and the
non-invertible nature of the Bloch parameters $\theta(q_m)$ and
$\phi(q_m)$, the final probability distributions are determined
numerically {\color{blue} without approximation.}
For a fixed rotation angle $\theta_r$, the two-mode state after Fock
damping and the rotation on mode~1, and before measurement, is denoted
$\ket{\Psi_2}$.  A $q$-homodyne measurement on mode~1 with outcome $q_m$
then occurs with probability density
\begin{equation}
  P(q_m)   
  = \frac{\big\|(\bra{q_m}\otimes \mathbb{1})\ket{\Psi_2}\big\|^2}
         {\langle\Psi_2|\Psi_2\rangle}.
  \label{eq:P_compact}
\end{equation}
The induced distributions on the Bloch sphere are obtained by
grid-based numerical pushforward method~\cite{Evans2005}.  For each
value of $\beta$ the symmetric homodyne window
$[q_{\min},q_{\max}]$ is adapted so that the cumulative probability
inside the window satisfies
$ \int_{|q_m|<q_{\max}} P(q_m)\, \mathrm{d}q_m
\;\gtrsim\; 1 - 10^{-7}$;
this guarantees that the tail probability outside the window is
$P_{\rm tail}\lesssim 10^{-7}$.  The interval $[q_{\min},q_{\max}]$ is
then discretized on a dense uniform grid with spacing
$\Delta q = (\sqrt{\pi})/\text{q\_dens}$, with typically
$\text{q\_dens}=10^5$ points per unit $\sqrt{\pi}$.  At each grid point the corresponding Bloch parameters are obtained numerically via
\begin{eqnarray}
  \phi &=& \tan^{-1}\!\left(\frac{{\rm Im}\,C_-}{{\rm Re}\,C_-}\right)
           -\tan^{-1}\!\left(\frac{{\rm Im}\,C_+}{{\rm Re}\,C_+}\right),
           \nonumber\\
  \theta &=& 2\tan^{-1}\!\left(\frac{|C_-|}{|C_+|}\right),
  \label{eq:Blochangles}
\end{eqnarray}
and a probability weight $P(q_{m,j})\,\Delta q$ is assigned to that
$(\theta,\phi)$ value.  Aggregating these weights over the grid yields
the marginal probability density functions $g(\theta)$ and $g(\phi)$, and
the two-dimensional Bloch-sphere PDFs are plotted in Fig.~2.

For fixed damping parameter $\beta$ and rotation angle $\theta_r$, the
homodyne outcome $q_m$ is, in principle, supported on the entire real
line.  The numerical analysis is performed using a finite symmetric window
$[q_{\min},q_{\max}] = [-q_{\max},q_{\max}]$ and a uniform grid
$q_{m,j} = -q_{\max} + j\,\Delta q$ with step size
$\Delta q$ (see below).  The corresponding discrete probabilities
$P(q_{m,j})\,\Delta q$ approximate the continuous density
$P(q_m)\,\mathrm{d}q_m$.

Rather than fixing $q_{\max}$ a priori, the window is chosen adaptively
for each $\beta$ so that the probability mass outside the window is
uniformly negligible.  Specifically, for each parameter set $(\beta,\theta_r)$, $q_{\max}$ is increased until the discrete cumulative
probability within the window satisfies
\begin{equation}
\int_{|q_m|<q_{\max}} P(q_m)\, \mathrm{d}q_m
\;\gtrsim\; 1 - 10^{-7}
\label{eq:tail_criterion}
\end{equation}
This guarantees that the tail probability
\begin{equation}
  P_{\text{tail}}
  = \int_{|q_m|>q_{\max}} P(q_m)\,\mathrm{d}q_m
\end{equation}
is bounded by $P_{\text{tail}}\lesssim 10^{-7}$ for all
$(\beta,\theta_r)$ used in the figures.

In practice, $q_{\max}$ depends primarily on $\beta$ (through
the overall width of the distribution).
Once a conservative window satisfying Eq.~\eqref{eq:tail_criterion} is determined for a given $\beta$, that window is fixed for all
$\theta_r$ at the same $\beta$. Representative values are
\begin{align}
  \beta = 0.04 &:\quad
  [q_{\min},q_{\max}] \approx [-20\sqrt{\pi},\,20\sqrt{\pi}], \\
  \beta = 0.001 &:\quad
  [q_{\min},q_{\max}] \approx [-120\sqrt{\pi},\,120\sqrt{\pi}],
\end{align}
in both cases satisfying $P_{\text{tail}} \lesssim 10^{-7}$.

\subsection{Grid resolution in $q_m$}

We parameterize the grid resolution by a density
\[
  \text{q\_dens} := \frac{1}{\Delta q}\;\;\text{(points per unit } \sqrt{\pi}\text{)},
\]
so that $\Delta q = (\sqrt{\pi})/\text{q\_dens}$.  Unless otherwise
stated, all main-text results use $\text{q\_dens}=100{,}000$, which
corresponds to $\Delta q \approx 3.2\times 10^{-5}\sqrt{\pi}$.

To verify that this discretization is sufficient, convergence checks are performed for representative parameter choices in both the
moderately damped and strongly damped regimes. As a diagnostic, the success probability is used
$P(F\ge F_0)$ of obtaining an output state with fidelity
$F\ge F_0$ to one of the four Pauli-related magic states, with a
threshold $F_0=0.96$.

\subsubsection{Lower squeezing regime: $\beta=0.04$}

For $\beta=0.04$ and $\theta_r = 0.062\pi$ ,the following values
of $P(F\ge 0.96)$ as a function of $\text{q\_dens}$ are obtained:
\begin{equation}
\begin{array}{c|c}
  \text{q\_dens} & P(F\ge 0.96) \\
  \hline
  10{,}000      & 0.38936425216436615 \\
  30{,}000      & 0.38934146623980910 \\
  100{,}000     & 0.38933827051966047 \\
  300{,}000     & 0.38933396944982807 \\
  1{,}000{,}000 & 0.38932854022624835 \\
  3{,}000{,}000 & 0.38932918743041920 \\
  10{,}000{,}000 & 0.38932898550358597
\end{array}
\label{tab:beta004}
\end{equation}
The value used in the main simulations,
$\text{q\_dens}=100{,}000$, differs from the finest grid,
$\text{q\_dens}=10^7$, by
\[
  |P_{10^5} - P_{10^7}|
  \approx 9.3\times 10^{-6},
\]
corresponding to a relative change below $2.5\times 10^{-5}$.  Moreover,
the results saturate as the grid is refined beyond
$\text{q\_dens}\approx 10^6$, with
$|P_{3\times 10^6} - P_{10^7}|\sim 2\times 10^{-7}$.  The reported probabilities are therefore converged to at least four
significant digits in this regime.

\subsubsection{Higher squeezing regime: $\beta=0.001$}

The most demanding case from the point of view of numerical resolution
is the smallest damping considered, $\beta=0.001$, where the homodyne
window is widest and the interference fringes in $P(q_m)$ are the
finest.  For $\beta=0.001$ and $\theta_r = 0.38467\pi$ we obtain:
\begin{equation}
\begin{array}{c|c}
  \text{q\_dens} & P(F\ge 0.96) \\
  \hline
  1{,}000      & 0.40477563394289745 \\
  3{,}000      & 0.40495869865312620 \\
  10{,}000     & 0.40477732206680450 \\
  30{,}000     & 0.40478719731072670 \\
  100{,}000    & 0.40480210345645460 \\
  300{,}000    & 0.40479583053813367 \\
  1{,}000{,}000 & 0.40479314835215340
\end{array}
\label{tab:beta0001}
\end{equation}
Here the difference between $\text{q\_dens}=100{,}000$ and
$\text{q\_dens}=1{,}000{,}000$ is
\[
  |P_{10^5} - P_{10^6}|
  \approx 9.0\times 10^{-6},
\]
again corresponding to a relative change of order $2\times 10^{-5}$.  The
small non-monotonic variations at the $10^{-5}$ level reflect the
interplay between sampling of very fine interference fringes and finite
floating-point precision, and are well below the accuracy required for
the plots in the main text.

\subsection{Non-orthogonality of effective basis states}

Let $\ket{\Psi_2}$ denote the two-mode state after the Fock damping and
the rotation $R_{\theta_r}$ on mode~1, and \emph{before} the $q$-homodyne
measurement on that mode.  Projecting onto $\bra{q_m}$ on mode~1 yields
\[
  \bra{q_m}\Psi_2\rangle = C_+(q_m)\ket{\tilde{+}} + C_-(q_m)\ket{\tilde{-}},
\]

where $\ket{\tilde{\pm}}$ are Fock-damped logical states.  In all numerical calculations, the exact expression is used for the homodyne
probability density,
\begin{equation}
P_q=\frac{||\langle q|\otimes \mathbb{1}\ket{\Psi_2}||^2}{\langle\Psi_2|\Psi_2\rangle}
  =2\frac{|C_+|^2\langle\tilde{+}|\tilde{+}\rangle
           + |C_-|^2\langle\tilde{-}|\tilde{-}\rangle
           + 2\operatorname{Re}[C_+^* C_-] \langle\tilde{+}|\tilde{-}\rangle}
         {(\langle\tilde{+}|\langle\tilde{0}|_r+\langle\tilde{-}|\langle\tilde{1}|_r)(|\tilde{+}\rangle|\tilde{0}\rangle_r+|\tilde{-}\rangle|\tilde{1}\rangle_r)},
\label{eq:P_exact}
\end{equation}
where $|\tilde{0}\rangle_r = R(\theta_r)\,|\tilde{0}\rangle$ and
$|\tilde{1}\rangle_r = R(\theta_r)\,|\tilde{1}\rangle$ denote the
Fock-damped logical GKP states on mode~1 after the rotation
$R(\theta_r)$ (and analogously for their bras).

No assumption is made that $\langle\tilde{+}|\tilde{-}\rangle=0$ at any stage but For completeness, the normalized overlap is also evaluated
\[
  \epsilon(\beta)
  = \frac{2\,\langle\tilde{+}|\tilde{-}\rangle}
         {\langle\tilde{+}|\tilde{+}\rangle
         +\langle\tilde{-}|\tilde{-}\rangle}
\]
for the range of $\beta$ used in the figures. This yields
\begin{equation}
\begin{array}{c|c}
  \beta & |\epsilon(\beta)| \\
  \hline
  0.10   & 7.56\times 10^{-4} \\
  0.04   & 5.88\times 10^{-9} \\
\end{array}
\end{equation}

For $\beta \le 0.04$, the true overlap is below numerical double-precision accuracy and is therefore reported as zero.

\medskip
In summary, the adaptive-window criterion
Eq.~\eqref{eq:tail_criterion}, the grid-resolution tests in
Tables~\eqref{tab:beta004}–\eqref{tab:beta0001}, show that truncation of the homodyne window,
discretization of $q_m$, do not affect the reported probability
distributions or fidelities at the level of precision used in this work.

\end{widetext}

\bibliographystyle{apsrev4-2}
\bibliography{refs}

\end{document}